\documentstyle[prl,twocolumn,aps,floats,epsfig]{revtex}

\begin{document}

\draft
\twocolumn[\hsize\textwidth\columnwidth\hsize\csname
@twocolumnfalse\endcsname

\title{The Mechanism of High-$T_{c}$ Superconductivity: ''Nonlinear''
Superconductivity}
\author{A. Mourachkine}
\address{Universit\'{e} Libre de Bruxelles, CP-232, 
Blvd du Triomphe, B-1050 Brussels, Belgium}

\maketitle

\vspace{2mm}
\begin{abstract}
\hspace{55mm} ''True laws of Nature cannot be linear.'' -  Albert Einstein
\vspace{3mm}

The main purpose of the paper is to present an overview of the current 
situation in the development of understanding of the mechanism of 
high-$T_{c}$ superconductivity which arises due to moderately strong, 
nonlinear electron-phonon interactions and due to magnetic 
(spin) fluctuations. The former are responsible for electron 
pairing, and the latter mediate the phase coherence.
\end{abstract}

\vspace{6mm}
]

\section{Introduction}

This Section is a {\em brief} reminder of key events which led
to understanding of the mechanism of high-$T_{c}$ superconductivity.

The first observation of what is now called the {\em soliton} was made
by John Scott Russell near Edinburgh (Scotland) in 1834. He was observing 
a boat moving on a shallow channel and noticed that, when the boat suddenly 
stopped, the wave that it was pushing at its prow ''rolled forward with 
great velocity, assuming the form of a large solitary elevation, a rounded, 
smooth and well defined heap of water which continued its course along 
the channel apparently without change of form or diminution of speed'' 
\cite{Russel}. He followed the wave along the channel for more than a 
mile.

The phenomenon of superconductivity was discovered by Dutch physicist 
H. Kamerlingh Onnes in 1911. He found that {\em dc} resistivity
of mercury suddenly drops to zero below 4.2 K \cite{Onnes}.

The microscopic theory of superconductivity in metals was proposed by 
J. Bardeen, L. Cooper and R. Schrieffer in 1957. The central 
concept of the BCS theory is weak electron-phonon interactions 
which lead to the appearance of an attractive potential between two 
electrons \cite{BCS}. 

To the best of my knowledge, the soliton (or bisoliton) model of 
superconductivity was for the first time considered by L. S. Brizhik and 
A. S. Davydov in 1984 \cite{bisoliton} in order to explain the 
superconductivity in organic quasi-one-dimensional (quasi-1D) 
conductors \cite{Jerome}. 

The interest in the research of superconductivity was renewed in 1986 
with the discovery of high-$T_{c}$ superconductivity in copper oxides 
(cuprates), made by J. G. Bednorz and K. A. M\"{u}ller \cite{cuprates}.

In 1987, L. P. Gor'kov and A. V. Sokol proposed the presence 
of two components of itinerant and more localized features in cuprates
\cite{Gorkov}. This kind of microscopic and dynamical phase separation
was later  rediscovered in other theoretical models.

By using the fact of the absence of the isotope effect in cuprates, which
is a hallmark of the BCS mechanism of superconductivity (in fact, 
there is the isotope effect in cuprates but it is very weak) A. S. Davydov 
proposed in 1988 the bisoliton mechanism of high-$T_{c}$
superconductivity \cite{Davydov1}.

The pseudogap above $T_{c}$ \cite{Phillips} was observed in 1989 in nuclear 
magnetic resonance (NMR) measurements \cite{Warren}.

In 1990, A. S. Davydov presented a theory 
of high-$T_{c}$ superconductivity, based on the concept of a moderately 
strong electron-phonon coupling which results in perturbation theory being 
invalid \cite{Davydov2,Davydov3}. The theory  utilizes the concept of
{\em bisolitons}, or electron (or hole) pairs coupled in a singlet state due to 
local deformation of the -O-Cu-O-Cu- chain in CuO$_{2}$ planes. We shall 
discuss the bisoliton model below.

In 1994, A. S. Alexandrov and N. F. Mott pointed out that, 
in cuprates, it is necessary to distinguish the ''internal'' wave function 
of a Cooper pair and the order parameter of the Bose-Einstein
condensate, which may have different symmetries \cite{Sasha}.

In 1995, V. J. Emery and S. A. Kivelson emphasized that superconductivity 
requires pairing and long-range phase coherence \cite{Emery}. In 
conventional superconductors described by the BCS theory, the pairing and 
the long-range phase coherence occur simultaneously at $T_{c}$ since the 
phase stiffness, which measures the ability of the superconducting state to 
carry supercurrent, is much larger than the energy gap, $\Delta$, which 
reflects the strength of the binding of electrons into Cooper pairs. In 
contrast to conventional superconductors, in cuprates, the energy gap and 
the phase stiffness have similar values \cite{Emery}. Therefore, the phase 
stiffness in cuprates is the weak link, and the pairing may occur above 
$T_{c}$ without the phase coherence which is established at $T_{c}$ 
\cite{Emery}.

In the same year 1995, J. M. Tranquada and co-workers \cite{Tranquada} 
found the presence of coupled, dynamical modulations of charges (holes) 
and spins in Nd-doped La$_{2-x}$Sr$_{x}$CuO$_{4}$ (LSCO) from neutron 
diffraction. In LSCO, antiferromagnetic stripes of copper spins are 
separated by 
periodically spaced quasi-1D domain walls to which the holes segregate. 
The spin direction in antiferromagnetic domains rotates by 180 degrees 
on crossing a domain wall. 

In 1997, V. J. Emery, S. A. Kivelson and O. Zachar presented the model 
of high-$T_{c}$ superconductivity based on the presence of charge stripes 
in CuO$_{2}$ planes \cite{Emery2}. They assumed that charge stripes are 
metallic, and there is the local separation of spin and charge along an 
individual stripe. Neutral spinons on a stripe acquire a spin gap via pair 
hopping between the stripe and its environment creating a singlet bound 
state. The phase coherence is established due to the Josephson coupling 
between stripes. It turned out that the model is incorrect, 
however, it is the first model of high-$T_{c}$ superconductivity based on 
the presence of charge stripes in CuO$_{2}$ planes. 

In 1999, on the basis of tunneling neutron scattering measurements, it 
was found that, in Bi$_{2}$Sr$_{2}$CaCu$_{2}$O$_{8+x}$ (Bi2212) and 
YBa$_{2}$Cu$_{3}$O$_{6+x}$ (YBCO), the phase coherence is established 
due to spin excitations \cite{AMour1,AMour9} which cause the appearance 
of the so-called magnetic resonance peak in inelastic neutron scattering 
spectra \cite{Mignod}.

In 2001, tunneling measurements provided evidence that 
quasiparticle peaks in tunneling spectra obtained in Bi2212 are caused 
by condensed solitonlike excitations, and the Cooper pairs in Bi2212 
seem to be Davydov's bisolitons \cite{AMour2,AMour3}. 
The data are discussed below.

\section{Nonlinear excitations: solitons}

For a long time linear equations have been used for describing different
phenomena. For example, Newton, Maxwell and Schr\"{o}dinger's
equations are linear, and they take into account only a linear response of
a system to an external disturbance. However, the majority of real systems 
are {\em nonlinear}. Most of the theoretical models are still relaying on a
{\em linear} description, corrected as much as possible for 
nonlinearities which are treated as small perturbations. It is well known 
that such approach can be absolutely wrong. The linear approach can 
sometimes miss completely some essential behaviors of the system. 

The solitary wave observed by J. S. Russell in 1834 on the water surface
is the nonlinear excitation called the soliton. Such waves can not be 
described by using linear equations. Unlike ordinary waves which represent 
a spatial periodical repetition of elevations and hollows on a water surface, 
or condensations and rarefactions of a density, or deviations from a mean 
value of various physical quantities, solitons are single elevations, such 
as thickenings {\em etc.}, which propagate as a unique entity with a given 
velocity. The transformation and motion of solitons are described by
nonlinear equations of mathematical physics.

Let us explore the world of solitons. 
A soliton is the extremely robust, nonlinear excitation 
localized in space, which has particlelike properties. There are three basic 
categories of solitons: the Korteweg-de Vries solitons (Russell's solitons), 
the Frenkel-Kontorova solitons and the envelope (group) solitons 
\cite{Filippov}. (The denomination ''the Frenkel-Kontorova solitons'' is used
here as a more general term ''the topological solitons''.)
The equation established 
by D. J. Korteweg and G. de Vries in 1895 describes Russell's solitons which
propagate at constant speed. The Frenkel-Kontorova solitons which can be 
moving or entirely static are described by the sine-Gordon equation. 

Yu. I. Frenkel and T. A. Kontorova theoretically predicted in 1939 solitons in 
chains of atoms and studied their properties \cite{Frenkel}, however, their 
real significance as well as their relation to the Russell soliton remained 
unknown for almost 30 years \cite{Filippov}. The Frenkel-Kontorova soliton 
is simpler than the Russell soliton. It is encountered in diverse physical 
systems. The Frenkel-Kontorova soliton has a fixed shape that does not 
depend on its velocity. It is interesting that the dependence of the energy 
$E$ of a Frenkel-Kontorova soliton on its velocity $v$ has the same form 
as for a relativistic particle
	\[ E = \frac{mv^{2}}{\sqrt{1-\frac{v^{2}}{v_{o}^{2}}}} \, ,\]
where $m$ is the mass of the Frenkel-Kontorova soliton, and $v_{o}$ is the 
longitudinal sound velocity in the chain. Thus, the Frenkel-Kontorova 
solitons behave like relativistic particles, and they cannot propagate faster 
than $v_{o}$. In addition, there also exist {\em antisolitons} which are 
analogous to antiparticles.

Solitons can also be found in the superconducting state. Vortices in type-II 
superconductors, which appear in the mixed state (Shubnikov's phase), are 
solitons. Frenkel-Kontorova solitons exist in long Josephson junctions, 
called Josephson solitons. A long Josephson junction is analogous to the 
chain of atoms studied by Frenkel and Kontorova \cite{Filippov}.

Solitons are encountered in biological systems in which the nonlinear
effects are often the predominant ones \cite{Davydov1,Davydov3,biology}. 
For example, many biological reactions would not occur without large 
conformational changes which cannot be described, even approximately, as 
a superposition of the normal modes of the linear theory. 

The shape of a nerve pulse was determined more than 100 years ago.
The nerve pulse has the bell-like shape and propagates with the velocity
of about 100 km/h (the diameter of nerves in mammals is less than 20 
microns) \cite{Filippov}. For almost a century, nobody realized 
that the nerve pulse is the soliton. So, all living creatures including humans 
are literally stuffed by solitons. Living organisms are mainly organic and, 
in principle, should be insulants. 

It is important to emphasize that almost all solitons described above 
are one-dimensional objects. For example, the Russell soliton can be 
approximately regarded as one-dimensional. There are only a few 
three-dimensional solitons known today, {\em e.g.} vortices and tornados
(in fact, tornados are a bound state of a few solitons spiralling around 
each other). Two-dimensional solitons are not observed experimentally, 
however, were discussed in a few theoretical papers \cite{Filippov}.

\section{Quasi-1D organic conductors, DNA, nanotubes and cuprates}

One may wonder what can be common among quasi-1D organic conductors
\cite{Jerome}, DNA (deoxyribonucleic acid) \cite{Kasumov1}, nanotubes 
\cite{Kasumov2} and cuprates. 
{\em Slightly doped, they all become superconducting at low temperature}. 
(Superconductivity in DNA is, in fact, induced \cite{Kasumov1}, however, if
one can find a technique to dope a DNA, it is most likely that the DNA will 
superconduct at low temperature by itself \cite{Kasumov3}.)

The second feature which is common among quasi-1D organic conductors
\cite{Davydov1,Davydov3,Ness}, DNA \cite{biology} and nanotubes 
\cite{Chamon,Astahova} 
is that {\em they transfer electrons (carry current) by solitons}. This 
signifies that electron-phonon interactions in these compounds are 
relatively strong and nonlinear \cite{Davydov1,Davydov3}. 

In the new MgB$_{2}$ superconductor, strong and nonlinear electron-phonon 
interactions are responsible for the high value of $T_{c}$ \cite{MgB2-2}. 

From these analogies, one can suppose that there are strong, nonlinear 
electron-phonon interactions in cuprates, which may lead to the formation 
of solitons.

\section{Solitons in cuprates}

Unexpectedly, tunneling measurements performed on Bi2212 single 
crystals provided evidence for solitonlike excitations
in Bi2212, which form the superconducting condensate 
\cite{AMour2,AMour3}. The main point of the evidence is presented here. 
For more details, the reader is referred to Refs\cite{AMour2,AMour3}.

It was found that, in Bi2212, tunneling spectra measured below $T_{c}$ 
are composite: they consist of incoherent part from the pseudogap and 
coherent quasiparticle peaks. The pseudogap which is observed above and 
below $T_{c}$ is a normal-state gap (a charge gap). Tunneling 
$I(V)$ characteristics corresponding exclusively to the quasiparticle peaks 
differ from the characteristics of BCS-type superconductors, but are in 
excellent agreement with theoretical curves derived for a bound state of 
two solitons. 

Figure 1 shows two theoretical $I(V)$ characteristics. The first one which 
corresponds to tunneling between a normal metal and a conventional 
superconductor is predicted by the so-called BTK theory \cite{BTK}. In 
metallic superconductors, the BTK predictions are verified by tunneling 
experiments. The second curve shown in Fig.1 corresponds to a bound state 
of two quasi-1D solitons \cite{french}. The difference between the two 
curves shown in Fig.1 is striking. If, in a conventional 
superconductor, the asymptotics of the $I(V)$ curve are quasi-linear, and 
current increases as the bias increases, the asymptotics of the 
characteristic corresponding to a bound state of solitons are horizontal, 
constant! If it is difficult to confuse these two $I(V)$ curves, however, 
their conductance peaks look very similar! (not the backgrounds)

The tunneling measurements performed on underdoped and overdoped single
crystals of Bi2212 show that $I(V)$ characteristics of the
quasiparticle peaks are similar to the $I(V)$ curve shown in Fig.1(b). This 
fact indicates that, in Bi2212, there is (quasi-) one dimensionality, and the 
Copper pairs consist of solitons. 

The most natural interpretation of the presence of quasi-one
\begin{figure}[t]
\leftskip-10pt
\epsfxsize=1.0\columnwidth
\centerline{\epsffile{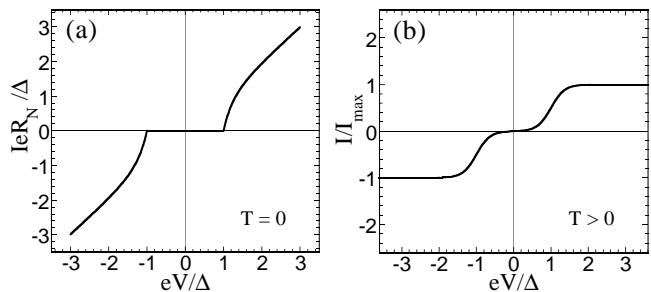}}
\vspace{2mm}
\caption{(a) Theoretical $I(V)$ characteristic for a junction between 
a conventional superconductor and a normal metal \protect\cite{BTK}. 
(b) Theoretical $I(V)$ characteristic of a bound state of two solitons 
\protect\cite{french}. }
\label{fig1}
\end{figure} 
dimensionality in Bi2212 is quasi-1D charge stripes \cite{Gorkov,Zaanen} 
which have been experimentally observed in cuprates, nickelates and 
manganites \cite{Serve}. Antiferromagnetic domains which separate the 
quasi-1D charge stripes are two dimensional \cite{Tranquada}. The 
solitons are quasi-1D objects, so they reside on dynamical charge stripes. 

Unfortunately, in cuprates, there is not much evidence for the presence 
of charge stripes below $T_{c}$. The reason is simple: it is not easy 
to detect them directly because they are truly dynamical. Almost all 
experimental techniques are {\em too slow} to observe the charge stripes 
at rest. The span time (or time of interaction) of a technique which is able 
to detect them has to be smaller than 1 ps = 10$^{-12}$ second. An 
illustration will make it clear. Watching a movie you are not able to realize 
that every second on the screen ''consists of'' 24 different exposures 
(snapshots) because the speed of the film is too fast for your eyes (in fact, 
for the brain) to see them separately.

Recently, solitons have been observed by tunneling spectroscopy on CuO 
chains in YBCO \cite{deLozanne}. The chains in YBCO are insulating, {\em i.e.} 
there are 2$k_{F}$ charge-density-waves (CDWs) along the chains 
\cite{Edwards}. The tunneling 
spectrum averaged along a CuO chain shows that, below $T_{c}$, there is a 
weak bound state of solitons inside the CDW gap \cite{note}. The magnitude
of {\em induced} superconducting gap is about 6 meV.

Solitons in cuprates are of the second type, {\em i.e.} they are the 
Frenkel-Kontorova (topological) solitons. Consequently, the 
superconductivity in cuprates is the relativistic phenomenon!

\section{The bisoliton model of superconductivity}

Here we shall briefly sketch Davydov's bisoliton model 
of superconductivity. For more details, the reader is referred to 
Refs\cite{Davydov2,Davydov3}.  

\subsection{Electron transfer by solitons}

First, we consider the electron transfer by solitons in molecular chains
(the main area of Davydov's research was the physics of biological 
processes).
Many biological phenomena, such as photosynthesis, relate to the electron
transfer from donor molecules to acceptor molecules through molecular 
structures. Davydov and co-workers theoretically showed that the
most energetically profitable transfer of electrons in protein molecules
occurs when there is a strong interaction of an electron with a chain. 
The system of an electron surrounded by local chain deformation is called 
an {\em electrosoliton}. When the static electrosoliton is generated, the 
total energy of the electron and the displacement field is decreased due 
to their interaction by the quantity 
\begin{equation}
\Delta E = m^{\ast}a^{2}\sigma^{4}/24k^{2}\hbar^{2},
\end{equation}
where $m^{\ast}$ is the mass of the electrosoliton (which exceeds the 
electron mass, $m$); $a$ is the distance 
between molecules along the chain; $\sigma$ is the deformation parameter 
of the chain, and $k$ is the coefficient of longitudinal elasticity of the chain. 
As $\Delta$$E$ increases, the electrosoliton becomes more stable. Large 
values of $\Delta$$E$, which provide the stability of the soliton under the 
conduction band of a quasiparticle (hole or electron), damps it strongly in 
the conduction band.

It is known that, in redox reactions occurring in living organisms, 
electrons are transferred from one molecule to another in pairs with 
opposite spins. It is also known that the transport of electrons in the 
synthesis process of ATP (adenosine triphosphate) molecules in conjugate 
membranes of mitochondria and chloroplasts is realized by pairs, but not 
individually. 

Two electrons coupled in a singlet state due to local chain deformation is
called a {\em bisoliton}. In the bisoliton state, both electrons move in the 
combined effective potential well
\begin{equation}
U_{\uparrow\downarrow}(\zeta) = -2g^{2}Jsech^{2}(g\zeta),
\end{equation}
relative to the coordinate frame $\zeta = (x-vt)/a$, where $x$ is the 
axis along the chain; $t$ is time, and $v$ is the velocity of the bisoliton. 
$J$ is the energy of the exchange interaction between neighboring 
molecules in the chain, and $g$ (and $G$) is the dimensionless quantity
\begin{equation}
g = m^{\ast}a^{2}G/\hbar^{2}, \; G = \sigma^{2}/k(1-s^{2}),
\end{equation}
where $s = v/v_{o}$ ($v_{o}$ is the longitudinal sound velocity in the
chain). The dimensionless parameter $g$ characterizes the coupling of an 
electron (hole) with the deformation field. The average distance between 
paired electrons in a bisoliton is determined by
\begin{equation}
L \approx 2\pi a /g.
\end{equation}
If $g \approx$ 1.5, then $L \approx 4a$.

The energy of pairing is given by
\begin{equation}
\Delta E(v) = \frac{\hbar^{2}g^{2}(1-5s^{2})}{4ma^{2}(1-s^{2})^{2}}\, ,
\end{equation}
where the effective Coulomb repulsion is neglected. 
At small electron velocities, a pairing becomes energetically profitable.
For this regime of velocities, the energy can be represented as follows
\begin{equation}
\Delta E(v) \approx ma^{2}\sigma^{4}(1-2s^{2})/4k^{2}\hbar^{2}, \; s^{2}\ll 1.
\end{equation}

The effective mass of paired electrons moving together with a local chain
deformation is given by
\begin{equation}
M_{bis} = 2m(1+2M\sigma^{4}/3k^{3}\hbar^{2}),
\end{equation}
where $M$ is the mass of a molecule. The mass of a bisoliton exceeds the
mass of two electrons.

If we take into account the Coulomb repulsion as a perturbation, then, at
small velocities, a pairing is still energetically profitable if the 
dimensionless coupling constant $g$ is greater than some critical value,
\begin{equation}
g_{cr} \equiv (4e_{eff}/bJ)^{1/2},
\end{equation}
where $e_{eff}$ is the effective screened charge, and $b$ is the average 
radius of molecules across the chain.

It is interesting to note that the pairing energy of a bisoliton, given by (6),
is independent of the mass $M$ of heavy molecules unlike that of the pairing
energy predicted by the BCS theory.

\subsection{Quasi-1D organic conductors}

The bisoliton model of superconductivity in quasi-1D organic conductors 
was proposed by Brizhik and Davydov \cite{bisoliton}. 

In quasi-1D organic conductors \cite{Jerome},  the bonds between plane 
molecules in stacks refer to weak van der Waals forces. Hence, due to the 
deformation interaction, a quasiparticle (electron or hole) causes a 
local deformation of a stack of molecules, which also induces
intramolecular atomic displacements in molecules. The deformation 
interaction results in the nonlinear equations, which were considered 
above.

If the density of excess pairs of quasiparticles in a stack is small, and the
condition $g <$ 3 takes place, the crystal becomes superconducting as the 
temperature decreases. The estimated critical temperature $T_{c}$, given by
\begin{equation}
k_{B}T_{c} \approx \sigma^{4}/2J^{2}k^{2}a^{4},
\end{equation}
is independent of the mass $M$ of heavy molecules. $k_{B}$ is the 
Boltzmann constant.

The paired quasiparticles are not localized between the molecules of 
stacks, but, surrounded by deformation, enveloping some molecules of 
stacks, move coherently as a unique entity along stacks of organic 
molecules without resistance. 

\subsection{Superconductivity in cuprates}

Davydov has applied the bisoliton model of superconductivity to 
high-$T_{c}$ superconductivity in cuprates. At that time, Davydov did not 
know about charge stripes which may exist in CuO$_{2}$ planes. 
He needed to locate one dimensionality in CuO$_{2}$ planes. The CuO$_{2}$ 
planes in cuprates consist of quasi-infinite parallel chains of alternating 
ions of copper and oxygen. He assumed that each -Cu-O-Cu-O- 
chain in the CuO$_{2}$ planes can be considered as a quasi-1D system. 
Therefore, the current flows along parallel chains. He studied the charge 
migration in one of these chains.
 
Due to electron-phonon interactions of quasiparticles with the resulting
displacements of the positions of elementary cells, there appears local
chain deformations which, in turn, lead to a coupling of quasiparticles
into singlet spin pairs, {\em i.e.} into bisolitons propagating along chains 
with a constant velocity $v < v_{o}$. {\em Bisolitons do not interact with 
acoustic phonons} since that interaction is taken completely into account 
in the coupling of quasiparticles with a local deformation. Therefore, 
they do not radiate phonons. At low temperature, bisolitons are stable
if the gain in the binding energy under their coupling exceeds the screened
Coulomb repulsion of their charges. We further assume that this condition
holds. 

The calculations of the energy gap, $\Delta$, in the quasiparticle spectrum 
resulting from a pairing give

\begin{equation}
2\Delta = g^{2}JF(q), 
\end{equation}
where the function $F(q)$ is determined by the modulus of the Jacobian 
elliptic function via the relation
\begin{equation}
F(q) = \frac{2}{E^{3}(q)}[2-q^{2}-\Xi (q)E^{-1}(q)], 
\end{equation}
where 
\begin{equation}
\Xi (q) = \frac{1}{3}[2(2-q^{2})E(q)-(1-q^{2})K(q)]. 
\end{equation}
The pairing gap is a half of the energy of the formation of a 
bisoliton. The functions $K(q)$ and $E(q)$ are the complete elliptic integrals 
of the first and the second kind, respectively. The value of the modulus $q$ 
is determined by the product of the period $L/a$ with the dimensionless 
coupling parameter $g$ and the elliptic integrals as follows
\begin{equation}
gL/a = 2E(q)K(q).
\end{equation}

The energy of pairing in (10) does not depend on the mass $M$ of an 
elementary cell. This mass only appears in the kinetic energy of the 
bisolitons. Therefore the isotope effect is absent, notwithstanding 
the fact that the basis of pairing is an electron-phonon interaction.

At rather small densities of quasiparticles, when the inequality 
$gL/a \gg 1$ is valid, the energy gap can be presented as
\begin{equation}
2\Delta = \frac{2}{3}g^{2}J[1+4g \frac{L}{a} exp(-gL/a)].
\end{equation}
At rather high density of quasiparticles, the energy gap which takes the 
value
\begin{equation}
2\Delta = 2g^{2}J\biggl[\frac{2}{\pi}\biggl]^{3}
\end{equation}
decreases in comparison with (14). In addition, Davydov did not take into 
account that $J$ (more precisely the effective exchange energy $J_{eff}$)
may depend on the doping level.

The permissible maximum density of quasiparticles (hole or electrons)
is determined by the minimum allowable distance $L_{min}$ between two 
bisolitons, given by
\begin{equation}
L_{min} = \pi^{2} a/2g. 
\end{equation}

In the bisoliton model of high-$T_{c}$ superconductivity, $T_{c}$ is 
defined as the temperature at which the energy gap vanishes. 

In superconductors described by the BCS theory, the dimensionless 
coupling parameter $g$ is the order of 10$^{-3}$--10$^{-4}$, while it is 
around 1 in cuprates. 

\section{The bisoliton model and experiment}

Here we discuss the bisoliton model of superconductivity and
compare data obtained in cuprates with predictions of the model. 

Davydov's genius is that, from only three experimental facts described 
below, he immediately understood that the high-$T_{c}$ superconductivity 
is related to soliton excitations. These three facts were: (i) 
the absence of the isotope effect (now we know that there is the isotope 
effect in cuprates but it is very weak \cite{Tallon}); (ii) the coherence 
length in hole-doped cuprates is very short, $\xi \approx$ 15--20 \AA, 
and (iii) cuprates become superconducting only when they are slightly 
doped. He concluded that the sharp decrease of the coherence length in 
cuprates in comparison with metallic superconductors indicates a rather 
large interaction of quasiparticles (holes or electrons) with the acoustic 
branch of the lattice vibrations inherent in cuprates.

In the bisoliton model, the mechanism of the establishment of phase 
coherence 
among bisolitons is not specified. In the framework of the bisoliton model, 
phonons cannot mediate the phase coherence because, as emphasized by 
Davydov, bisolitons do not interact with acoustic phonons. In the BCS 
theory for conventional superconductors, the phase coherence among the
Cooper pairs is established due to the overlap of their wave functions (the 
wave-function coupling) because the average distance between the Cooper 
pairs is much smaller than the coherence length (the size of a Cooper pair). 
In cuprates, the distance between the Cooper pairs (bisolitons) is similar to 
the size of the bisoliton. Under such conditions, can the wave-function 
coupling mediate the phase coherence among 
the bisolitons? Davydov defined $T_{c}$ as the temperature at which the 
energy gap vanishes. With such a definition for $T_{c}$, the wave-function 
coupling can formally be the mediator of the phase coherence. 
However, in reality, the wave-function coupling cannot be responsible for 
mediating the phase coherence among the bisolitons. This was the reason 
why Davydov avoided to discuss the bell-like shape of $T_{c}(p)$ dependence
(see Fig.3), where $p$ is the hole (electron) concentration in CuO$_{2}$ 
planes, because he could not explain it in the framework of the model. The 
bisoliton theory predicts that by increasing the hole (electron) 
concentration the magnitude of pairing gap decreases. Consequently, if 
the wave-function coupling mediate the phase coherence among the 
bisolitons, then, from $T_{c} \sim \Delta$, one can obtain that by 
increasing the hole (electron) concentration $T_{c}$ will monotonically 
decrease, contrary to the experiment. Thus, in cuprates, {\em the 
phase coherence among bisolitons is established due to a non-phonon 
mechanism which is different from the wave-function coupling}.
 
Experimentally, in hole-doped cuprates, spin fluctuations (magnetic 
electron-electron interactions) mediate the phase coherence among the Copper 
pairs \cite{AMour1,AMour9}. So, the bisoliton model is the theory of 
soliton pairing, but it lacks the mechanism of the establishment of phase 
coherence. Therefore, we now discuss solely the pairing characteristics. 

First, let us estimate the coupling constant $g$ in hole- and electron-doped
cuprates. By using the values of Cu-O-Cu bonding length 
($a \simeq$ 3.9 \AA) and the coherence length measured in 
hole-doped ($\xi \approx$ 15--20 \AA) and electron-doped 
($\xi \approx$ 80 \AA) cuprates, from (4), we have $g_{h} \simeq$ 
1.2--1.6 and $g_{e} \simeq$ 0.3, respectively. The ratio $g_{h}/g_{e} 
\simeq$ 4--5.3 is in good agreement with similar ratio estimated for 
C$_{60}$, $g_{h}/g_{e} \approx$ 5--6 \cite{Batlog}. This means that, 
independently from the material, the maximum $T_{c}$ value will 
always be higher in hole-doped superconductors.

Let us now estimate the maximum concentration of doped holes (electrons) 
in cuprates, $p_{max}$, at which the bisolitons still exist. From (4) and 
(16), one can easily obtain that, on a single chain (stripe {\it etc}.), 
	\[ p_{max} = \frac{4g}{\pi (\pi+4)} \, .\]
To estimate $p_{max}$ in hole-doped cuprates, let us use $g \simeq$ 1.5. 
Then, the calculations give $p_{max} \simeq$ 0.27 in hole-doped and 
$p_{max} \simeq$ 0.054 in electron-doped cuprates. The maximum 
hole/electron concentration at which hole- and electron-doped cuprates 
are still superconducting is 0.27 and 0.17, respectively. So, in hole-doped
cuprates, the prediction of the bisoliton model coincides with the real 
value. However, in electron-doped cuprates, it is three times smaller 
than the measured one. 

In (14) and (15), the energy gap $\Delta$ depends on the energy of the 
exchange interaction $J$, the hole (electron) concentration $p$ 
($\sim a/L$) and the coupling parameter of the electron-phonon 
interactions $g$. Both $J$ ({\em i.e.} $J_{eff}$) and $g$ depend themselves  
on the doping level. Since the coupling parameter depends weakly on $p$, 
the variations of the energy gap as a function of $p$ is mainly determined 
by the $J_{eff}(p)$ dependence. Raman scattering measurements in Bi2212 
show that, at different hole concentrations, $\Delta \approx 
\frac{2}{3}J_{eff}$ (see Fig.2 in Ref.\cite{Klein}). In order to estimate the 
energy gap from (14) and (15), let us use again $g \simeq$ 1.5. Then, at low 
hole concentrations, we have $\Delta \simeq \frac{g^{2}}{3}J_{eff} \simeq 
\frac{2}{3}J_{eff}$, and, at high hole density, $\Delta \simeq 
\frac{g^{2}}{4}J_{eff} \simeq \frac{1}{2}J_{eff}$. So, there is very good 
agreement between the bisoliton model and the data. 

To conclude, in hole-doped cuprates, the bisoliton model of 
superconductivity is correct in the description of pairing 
characteristics, however, it lacks the mechanism of the establishment of 
phase coherence. It is a pity that the bisoliton model did not attract any 
attention on earlier stages of the development of high-$T_{c}$ 
superconductivity. It is also a pity that A. S. Davydov is no longer with us 
\cite{Davydov4}.

\section{Charge stripes}
  
The origin of the driving force for the charge-stripe formation in cuprates
is still an open question. Since the charge stripes in cuprates appear
immediately after the lattice transformation \cite{Tranquada}, it is 
reasonable to assume that electron-lattice interactions are responsible
for the formation of charge stripes (see also Ref.\cite{Khomski}). In other 
words, solitons which later form the bisolitons reside on charge stripes 
not because the charge stripes existed beforehand, but because solitons 
and charge stripes appear simultaneously as a consequence of 
electron-lattice interactions. 

In fact, the formation of dynamical charge stripes in cuprates can 
be viewed as the appearance of dynamical CDWs \cite{CDW,Mark,Bianconi} 
in consequence of electron-phonon interactions.

\section{The mechanism of superconductivity in cuprates}

In this Section, I present a brief description of the mechanism of 
high-$T_{c}$ superconductivity. For more details, the reader is referred 
to Refs\cite{AMour2,AMour3}. 

The Cooper pairs in cuprates are bisolitons or hole (electron) pairs 
coupled in a singlet state due to local deformation of the lattice. Thus, 
moderately strong, nonlinear electron-phonon interactions are responsible 
for the pairing. The bisolitons are formed above $T_{c}$ on charge stripes 
(on dynamical CDWs). 

The long-range phase coherence among the bisolitons is established due to 
spin fluctuations in local antiferromagnetic domains of CuO$_{2}$ planes, 
which are situated between charge stripes. However, in different cuprates,
the phase coherence is established differently. 
Let us classify all hole-doped superconducting cuprates in the two 
categories: 40 K and 90 K cuprates. All optimally doped cuprates having the 
$T_{c}$ value less (more) than 40 K (90 K) belong to the first (second) group. 
{\it De fait}, the 40 K cuprates have one CuO$_{2}$ layer per unit cell. 
The main difference between the two groups, which defines the $T_{c}$ 
value, is the absence/presence of the so-called magnetic resonance peak in 
inelastic neutron scattering (INS) spectra. The phase 
coherence in 40 K cuprates is locked at $T_{c}$ via the long-range 
antiferromagnetic order (or spin-density-wave order). In 90 K cuprates, the 
phase coherence is established among bisolitons due to spin excitations 
which cause the appearance of the magnetic resonance peak in INS spectra. 
The spin excitations seem to be in resonance with the bisolitons.

Such a mechanism of the establishment of phase coherence implies that 
the magnetic resonance peak can be observed in all 90 K cuprates. 
The resonance peak has already been detected by INS in the double-layer 
cuprates YBCO and Bi2212 \cite{AMour1,AMour9} and in the single-layer 
Tl$_{2}$Ba$_{2}$CuO$_{6}$ \cite{Tl2201}. The process of bisoliton 
condensation at $T_{c}$, as a matter of fact, is the Bose-Einstein 
condensation. In cuprates, below $T_{c}$, charge, spin and lattice degrees of 
freedom in the CuO$_{2}$ planes are coupled.

The symmetry of the pairing wave function of bisolitons is an 
anisotropic s-wave 
while the order parameter which is responsible for the establishment of 
phase coherence has the $d_{x^{2}-y^{2}}$ (d-wave) symmetry. This is the 
reason why all phase-sensitive techniques detect in cuprates the d-wave 
symmetry. At the same time, tunneling measurements show a s-wave 
symmetry of the condensate \cite{Klemm}, {\em i.e.} the symmetry of the 
pairing wave function. The maximum magnitudes of the two gaps in Bi2212 
as functions of doping are schematically shown in Fig.2 as well as their 
temperature dependences.

In cuprates, above and below $T_{c}$, there is a normal-state gap, 
\begin{figure}[t]
\leftskip-10pt
\epsfxsize=0.9\columnwidth
\centerline{\epsffile{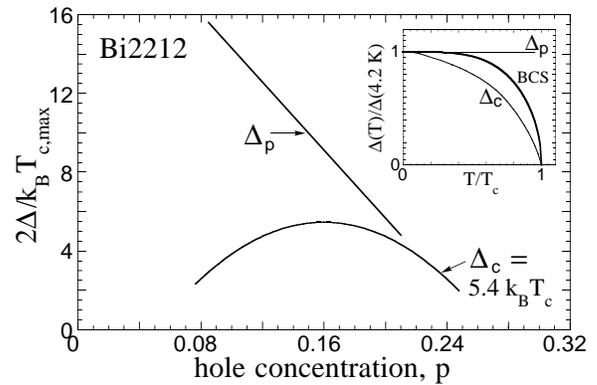}}
\vspace{2mm}
\caption{Phase diagram of Bi2212: $\Delta _{p}$ is the pairing gap of 
bisolitons, and $\Delta _{c}$ is the phase-coherence gap. $\Delta _{p}$
is caused by moderately strong, nonlinear electron-phonon interactions, 
whereas $\Delta _{c}$ has the magnetic origin due to spin fluctuations 
(excitations). The inset schematically shows the temperature dependences 
of $\Delta _{p}$ and $\Delta _{c}$ in the underdoped and optimally doped 
regions. The thick line corresponds to the BCS temperature 
dependence. (From Ref.\protect\cite{AMour3}). The phase diagram of other 
cuprates is similar to that of Bi2212.}
\label{fig2}
\end{figure}
a pseudogap. It is impossible to formulate the origin of 
the pseudogap because, in fact, there are two pseudogaps, and different 
techniques observe different pseudogaps. Besides the presence of the 
bisoliton pairing gap above $T_{c}$, there is a charge gap (a CDW gap) on 
charge stripes, and there is a spin gap in local antiferromagnetic domains 
situated between charge stripes \cite{Zaanen,Millis}. The 
superconducting cuprates inherited antiferromagnetic correlations from 
their parent compounds, antiferromagnetic Mott insulators. Figure 3 
schematically shows the characteristic temperatures 
of charge and spin degrees of freedom in CuO$_{2}$ planes of Bi2212.

In transport measurements, the pseudogap which relates to the spin gap 
\cite{Mochalk} vanishes at $p \simeq$ 0.19--0.2 \cite{Wuts} because the 
electronic state in CuO$_{2}$ planes above $p \simeq$ 0.19--0.2 becomes 
inhomogeneous in a macroscopic scale \cite{Wen}. In other words, in the 
heavily overdoped region, there exists already metallic islands in 
CuO$_{2}$ planes.

The scenario of superconductivity in electron-doped cuprates seems to be 
similar to the scenario realized in 40 K hole-doped cuprates.
\begin{figure}[t]
\leftskip-10pt
\epsfxsize=0.9\columnwidth
\centerline{\epsffile{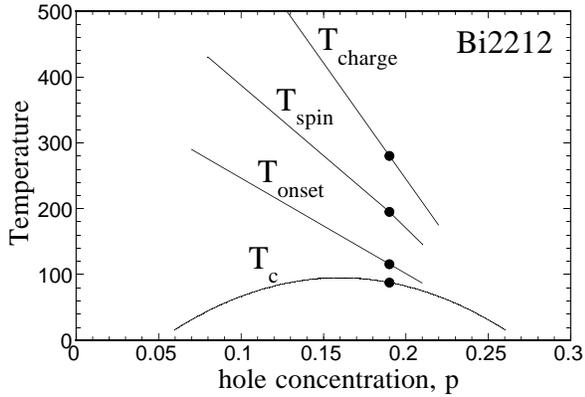}}
\vspace{2mm}
\caption{Characteristic temperatures in Bi2212: $T_{charge}$ is the 
temperature of the charge-stripe formation (and solitons and a CDW gap 
on charge stripes). $T_{spin}$ is the temperature of the formation of 
antiferromagnetic domains between charge stripes (and a spin gap). 
$T_{onset}$ is the onset of bisolitons on charge stripes (see also Fig.2). 
The long-range phase coherence among bisolitons appears at $T_{c}$. The 
circles are measured points from Refs\protect\cite{Ekino,AMour4} (see 
explanations in Ref.\protect\cite{AMour4}). The straight lines are shown 
schematically and, in reality, they are curves. $T_{charge}$, $T_{spin}$ 
and $T_{onset}$ are in-plane characteristics, whereas $T_{c}$ is 
{\em mainly} a $c$-axis characteristic.}
\label{fig3}
\end{figure}

Many data obtained in cuprates can be naturally understood in the 
framework of such a scenario for high-$T_{c}$ superconductivity, as
analyzed elsewhere \cite{AMour3}. As an example and 
additional evidence, let us consider elastic properties of high-$T_{c}$ 
superconductors, which are remarkably different from those of 
conventional superconductors. 
In cuprates, the elastic coefficients for various modes reveal not 
only anisotropic lattice properties in the normal state, but also anisotropic 
coupling between superconductivity and lattice deformation 
\cite{sound,sound1}. What is even more striking is the effect of magnetic 
field applied along various axes on the longitudinal elastic coefficients: 
they exhibit exclusively below $T_{c}$ a strong and anisotropic dependence 
on the direction and magnitude of the applied magnetic field \cite{sound}. 
All these unusual elastic properties of cuprates can naturally be understood 
in the framework of the high-$T_{c}$ superconductivity scenario described 
above. In contrast to characteristics of conventional superconductors, all 
bisoliton characteristics depend on the longitudinal elastic coefficients 
and longitudinal sound velocity (see Section V). Second, since the bisolitons 
are aligned along -Cu-O-Cu-O- 
chains in CuO$_{2}$ planes \cite{Davydov1,Davydov2}, it is {\em expected} 
that the elastic coefficients for different modes exhibit anisotropic lattice 
properties below and somewhat above $T_{c}$. Third, the mysterious 
behavior of longitudinal elastic coefficients seen in magnetic fields 
below $T_{c}$ \cite{sound} takes place because the spin fluctuations 
(excitations) are coupled below $T_{c}$ to the bisolitons and, thus, to the 
lattice. Consequently, the lattice is affected by the applied magnetic field 
through the intermediary of spin fluctuations. Finally, a hardening of the
longitudinal elastic coefficient at very low temperature occurs due to 
the formation of spin glass.

So, the general belief which was predominant for the last 14 years, 
namely, that the superconductivity in cuprates has nothing to do with 
electron-phonon interactions, is wrong. On the contrary, the 
superconductivity in cuprates occurs due to electron-phonon interactions 
which are moderately strong and nonlinear (and with assistance of spin 
fluctuations).

There are two types of electron pairing known today: "linear" (the BCS type) 
and "non-linear".

\section{Key Experiments for bisoliton superconductivity}

The bisoliton superconductivity requires moderately strong electron-phonon 
interactions, thus, it cannot occur in pure metals at {\em normal} conditions.
The first indication of bisoliton superconductivity is the coherence length. 
As emphasized by Davydov, the short coherence length, of say less than 
100 \AA, can already be a good sign of bisoliton superconductivity. 

By analogy with the isotope effect in conventional superconductors, the 
study of elastic properties in cuprates and other compounds can serve as 
a key experiment for revealing the bisoliton superconductivity. 

In addition, tunneling measurements always remain a key 
experiment for any type of superconductivity \cite{AMour2,Eric}. 

An attribute of the magnetic (spin-fluctuation) mechanism is the 
temperature dependence of the gap, lying below the BCS curve, as shown 
in the inset of Fig.2.

\section{Superconductivity at 300 K}

Due to a clear picture of the mechanism of high-$T_{c}$ 
superconductivity, it possible to discuss necessary ''ingredients'' of
superconductivity at high temperature. This discussion does not 
signify that all types of superconductivity which will be discovered in 
the future have to be based on electron-phonon interactions. However, by 
taking into account the importance of electron-phonon interactions in the 
BCS mechanism of superconductivity and in the bisoliton model of 
high-$T_{c}$ superconductivity it is clear that the electron-phonon 
interactions will surprise us repeatedly in the future.

Let us discuss necessary ingredients of superconductivity at high 
temperature ($>$ 200 K) by using the scenario of superconductivity in 
cuprates as a basis. First, the system has to be hole-doped and quasi-1D.
It is important that the system is {\em quasi}-1D because one-dimensional 
systems tend not to be superconducting by themselves \cite{Emery2} like, 
for example, the chains in YBCO. Second, a rather large interaction of holes 
with the acoustic branch of lattice vibrations is necessary for pairing at high 
temperature. Third, the presence of strong magnetic (spin) fluctuations is 
vital to mediate the phase coherence and, 
in each system, it is necessary to understand the best conditions for the 
appearance of the resonance mode. The latter ingredient, probably, will be 
always the weakest link, the bottleneck in achieving high $T_{c}$. 

Superconductivity at 300 K, in principle, is possible. As a matter of fact, 
bisolitons (the Cooper pairs) exist above 300 K in living tissues 
\cite{Davydov1,Davydov3}. The question is how to make them to 
communicate with each other? 

\acknowledgments
I thank A. M. Gabovich for drawing my attention to the works 
of A. S. Davydov, and I would like to thank A. Yu. Kasumov, Ph. Bourges, 
A. V. Buryak and G. Gusman for discussions. 


\vspace{-5mm}
\begin{references}
\vspace{-15mm}

\bibitem{Russel} J. S. Russell, {\it ''Report on Waves''}, in {\it Rep. 14$^{th}$ 
Meet. British Assoc. Adv. Sci.} (John Murray, 1844) 311.

\bibitem{Onnes} H. Kamerlingh Onnes, {\it Commun. Phys. Lab. Univ. Leiden} 
{\bf 124c} (1911).

\bibitem{BCS} J. Bardeen, L. N. Cooper, and J. R. Schrieffer, {\it Phys. Rev.}
{\bf 108}, 1175 (1957). 

\bibitem{bisoliton} L. S. Brizhik and A. S. Davydov, {\it Fiz. Nizk. 
Temperatur}, {\bf 10} 358 (1984).

\bibitem{Jerome} D. J\'{e}rome, A. Mazaud, M. Ribault, and K. Bechgaard, 
{\it J. Phys. (Paris) Lett.} {\bf 41}, L95 (1980).

\bibitem{cuprates} J. G. Bednorz and K. A. M\"{u}ller, {\it Z. Phys. B} 
{\bf 64}, 189 (1986).

\bibitem{Gorkov} L. P. Gor'kov and A. V. Sokol, {\it JETP Lett.} {\bf 46},
420 (1987).

\bibitem{Davydov1} A. S. Davydov, {\it Solitons in Molecular Systems} 
(Naukova Dumka, Kiev, 1988), in Russian.

\bibitem{Phillips} J. C. Phillips, {\it Phys. Rev. Lett.} {\bf 59}, 1856 (1987).

\bibitem{Warren} W. W. Warren, Jr., R. E. Walstedt, G. F. Brennert, R. J. Cava, 
R. Tycko, R. F. Bell, and G. Dabbagh, {\it Phys. Rev. Lett.} {\bf 62}, 1193 (1989).

\bibitem{Davydov2} A. S. Davydov, {\it Phys. Rep.} {\bf 190}, 191 (1990). 

\bibitem{Davydov3} A. S. Davydov, {\it Solitons in Molecular Systems} 
(Kluwer Academic, Dordrecht, 1991).

\bibitem{Sasha} A. S. Alexandrov and N. F. Mott, {\it Rep. Prog. Phys.} {\bf 57}, 
1197 (1994).

\bibitem{Emery} V. Emery and S. Kivelson, {\it Nature} {\bf 374}, 434 (1995).

\bibitem{Tranquada} J. M. Tranquada, B. J. Sternlieb, J. D. Axe, Y. Nakamura,
and S. Uchida, {\it Nature} {\bf 375}, 561 (1995).

\bibitem{Emery2} V. Emery, S. Kivelson, and O. Zachar, {\it Phys. Rev. B}
{\bf 56}, 6120 (1997).

\bibitem{AMour1} A. Mourachkine, {\it J. Low Temp. Phys.} {\bf 117}, 401 
(1999), and references therein.

\bibitem{AMour9} A. Mourachkine, {\it Europhys. Lett.} {\bf 55}, 86 (2001).

\bibitem{Mignod} J. Rossat-Mignod, L. P. Regnault, C. Vettier, Ph. Bourges,
P. Burlet, J. Bossy, J. Y. Henry, and G. Lapertot, {\it Physica C}
{\bf 185-189}, 86 (1991). 

\bibitem{AMour2} A. Mourachkine, {\it Europhys. Lett.} {\bf 55}, 559 
(2001).

\bibitem{AMour3} A. Mourachkine, {\it Supercond. Sci. Technol.} {\bf 14}, 
329 (2001).

\bibitem{Filippov} A. T. Filippov, {\it The Versatile Soliton} (Birkh\"{a}user,
Boston, 2000).

\bibitem{Frenkel} Yu. I. Frenkel and T. A. Kontorova, {\it J. Phys. Moscow} 
{\bf 1}, 137 (1939).

\bibitem{biology} M. Peyrard, ed., {\it Nonlinear Excitations in Biomolecules}
(Springer-Verlag, Berlin, 1995).

\bibitem{Kasumov1} A. Yu. Kasumov, M. Kaciak, S. Gu\'{e}ron, B. Reulet, 
V. T. Volkov, D. V. Klinov, and H. Bouchiat, {\it Science} {\bf 291}, 280 (2001).

\bibitem{Kasumov2} M. Kaciak, A. Yu. Kasumov, S. Gu\'{e}ron, B. Reulet, 
I. I. Khodos, Yu. B. Gorbatov, V. T. Volkov, L. Vaccarini, and H. Bouchiat, 
{\it Phys. Rev. Lett.} {\bf 86}, 2416 (2001).

\bibitem{Kasumov3} A. Yu. Kasumov, private communications.

\bibitem{Ness} H. Ness, S. A. Shevin, and A. J. Fisher, {\it Phys. Rev. B}
{\bf 63}, 125422 (2001).

\bibitem{Chamon} C. Chamon, {\it Phys. Rev. B} {\bf 62}, 2806 (2000).

\bibitem{Astahova} M. T. Figge, M. Mostovoy, and J. Knoester, 
{\it Phys. Rev. Lett.} {\bf 86}, 4572 (2001).

\bibitem{MgB2-2} T. Yildirim, O. Gulseren, J. W. Lynn, C. M. Brown, 
T. J. Udovic, H. Z. Qing, N. Rogado, K. A. Regan, M. A. Hayward, J. S. Slusky, 
T. He, M. K. Haas, P. Khalifah, K. Inumaru, and R. J. Cava, cond-mat/0103469.

\bibitem{BTK} G. E. Blonder, M. Tinkham and T. M. Klapwijk, {\it Phys. Rev. B} 
{\bf 25}, 4515 (1982).

\bibitem{french} M. Remoissenet, {\it Waves Called Solitons} 
(Springer-Verlag, Berlin, 1999).

\bibitem{Zaanen} J. Zaanen and O. Gunnarsson, {\it Phys. Rev. B} {\bf 40},
7391 (1989).

\bibitem{Serve} B. Goss Levi, {\it Physics Today}, issue {\bf 51/6}, 19
(1998).

\bibitem{deLozanne} A. L. de Lozanne, D. J. Derro, E. W. Hudson, K. M. Lang,
S. H. Pan, J. C. Davis, and J. T. Markert, oral presentation at the
"Stripes 2000" Conference in Rome, 25-30 September 2000.

\bibitem{Edwards} H. L. Edwards, A. L. Barr, J. T. Markert, and 
A. L. de Lozanne, {\it Phys. Rev. Lett.} {\bf 73}, 1154 (1994).

\bibitem{note} Note that, in Ref.\cite{deLozanne}, the data were 
presented without interpretation. 

\bibitem{Tallon} D. J. Pringle, G. V. M. Williams, and J. L. Tallon,
{\it Phys. Rev. B} {\bf 62}, 12527 (2000).

\bibitem{Batlog} J. H. Sch\"{o}n, Ch. Kloc, and B. Batlogg. {\it Nature} 
{\bf 549}, 549 (2000).

\bibitem{Klein} M. V. Klein, {\it Physica C} {\bf 341-348}, 2173 (2000).

\bibitem{Davydov4} A. S. Davydov, 1912-1993; 
http://www.bitp.kiev/bitp /htmls/dav.htm

\bibitem{Khomski} D. I. Khomskii and K. I. Kugel, {\it Europhys. Lett.} 
{\bf 55}, 208 (2001). 

\bibitem{CDW} C. Nayak and F. Wilczek, {\it Phys. Rev. Lett.} {\bf 78},
2465 (1997).

\bibitem{Mark} R. S. Markiewicz, {\it Phys. Rev. B} {\bf 62}, 1252 (2000).

\bibitem{Bianconi} A. Bianconi, D. Di Castro, G. Bianconi, 
A. Pifferi, N. L. Saini, F. C. Chou, D. C. Jonston, and M. Colapietro, 
{\it Physica C} {\bf 341-348}, 1719 (2000).

\bibitem{Tl2201} Ph. Bourges, private communications.

\bibitem{Klemm} See, for example, Q. Li, Y. N. Tsay, M. Suenaga, R. A. Klemm
G. D. Gu, and N. Koshizuka, {\it Phys. Rev. Lett.} {\bf 83}, 4160 (1999).

\bibitem{Millis} A. J. Millis and H. Monien, {\it Phys. Rev. Lett.} 
{\bf 70},  2810 (1993).

\bibitem{Mochalk} V. V. Moshchalkov, L. Trappeniers, and J. Vanacken,
{\it Europhys. Lett.} {\bf 46}, 75 (1999).

\bibitem{Wuts} B. Wuyts, V. V. Moshchalkov, and Y. Bruynseraede, 
{\it Phys. Rev. B} {\bf 53}, 9418 (1996).

\bibitem{Wen} H. H. Wen, W. L. Yang, and Z. X. Zhao, {\it Physica C} 
{\bf 341-348}, 1735 (2000).

\bibitem{sound} T. Fujita and T. Suzuki, in {\it Physics of High-Temperature
Superconductors}, edited by S. Maekawa and M. Sato (Springer-Verlag, Berlin,
1992), p. 333. 

\bibitem{sound1} M. Lei, A. Migliori, and H. Ledbetter, {\it Physica C} 
{\bf 282-287}, 1077 (1997).


\bibitem{Eric} E. W. Hudson, K. M. Lang, V. Madhavan, S. H. Pan, H. Eisaki,
S. Uchida, and J. C. Davis, {\it Nature} {\bf 411}, 920 (2001).

\bibitem{Ekino} T. Ekino, Y. Sezaki, and H. Fujii, {\it Phys. Rev. B} {\bf 60}, 
6916 (1999).

\bibitem{AMour4} A. Mourachkine, {\it Europhys. Lett.} {\bf 49}, 86 (2000).

\end{references}
\end{document}